\documentclass[preprint,showpacs,superscriptaddress,preprintnumbers,longbibliography,amsmath,amssymb,aps,pre,floatfix,letter]{revtex4}

\usepackage{graphicx}
\usepackage{dcolumn}
\usepackage{bm}
\usepackage{amsmath}
\usepackage{color} 
\usepackage{pifont} 
\usepackage{stmaryrd} 
\usepackage{float}

\def\camII@III{II/III}
\def\camIV{IVC$\beta$}
\def\camVI{VI}
\def\camLGN{LGN}
\def\mV{\ensuremath{~\textnormal{mV}}}
\def\Sps{\ensuremath{\mathcal{S}}}
\def\Gsc{\ensuremath{\mathcal{G}}}
\def\Hsc{\ensuremath{\mathcal{H}}}
\def\Pdist{\ensuremath{\mathcal{P}}}
\def\Fdist{\ensuremath{\mathcal{F}}}
\def\sabs{\ensuremath{\bar{s}}}
\def\aabs{\ensuremath{\bar{\alpha}}}
\def\Dabs{\ensuremath{\bar{D}}}
\def\Tabs{\ensuremath{\bar{T}}}
\def\tabs{\ensuremath{\bar{\tau}}}
\def\muabs{\ensuremath{\bar{\mu}}}

\def\virgula{\ \textnormal{,}} 

\definecolor{red1}{RGB}{249,42,42}
\definecolor{red2}{RGB}{194,29,29}
\definecolor{red3}{RGB}{146,18,18}
\definecolor{red4}{RGB}{89,5,5}
\definecolor{blue1}{RGB}{42,42,249}
\definecolor{blue2}{RGB}{29,29,194}
\definecolor{blue3}{RGB}{18,18,146}
\definecolor{blue4}{RGB}{5,5,89}
\definecolor{gray1}{rgb}{0.9,0.9,0.9}
\definecolor{gray2}{rgb}{0.8,0.8,0.8}
\definecolor{redC}{RGB}{210,45,45}
\definecolor{blueC}{RGB}{112,179,250}
\definecolor{greenC}{RGB}{99,199,113}
\definecolor{purpleC}{RGB}{194,129,201}

\makeatletter
\def\ifmyemptyarg#1{%
  \if\relax\detokenize{#1}\relax 
    \expandafter\@firstoftwo
  \else
    \expandafter\@secondoftwo
  \fi}
\makeatother
\newcommand*{\autorano}[2][]{%
    \ifmyemptyarg{#1}{%
        \citeauthor{#2}~\cite{#2}%
    }{%
        \citeauthor{#2}~\cite[#1]{#2}%
    }%
} 

\begin{document}

\preprint{?APS/123-QED?}

\title{Measuring neuronal avalanches in disordered systems with absorbing states}

\author{M.~Girardi-Schappo}
\affiliation{
Neuroimaging of Epilepsy Laboratory, McConnell Brain Imaging Center, McGill University, Montreal Neurological Institute and Hospital,
H3A 2B4, Montreal, Quebec, Canada
}
\affiliation{
Departamento de F{\'i}sica, Universidade Federal de Santa Catarina, 88040-900, Florian{\'o}polis, Santa Catarina, Brazil
}




\author{M.~H.~R.~Tragtenberg}
\email{marcelotragtenberg@gmail.com}
\affiliation{
Departamento de F{\'i}sica, Universidade Federal de Santa Catarina, 88040-900, Florian{\'o}polis, Santa Catarina, Brazil
}

\date{\today}

\begin{abstract}
Power-law shaped avalanche size distributions are widely used to probe for critical behavior in many different systems, particularly in
neural networks. The definition of avalanche is ambiguous.
Usually, theoretical avalanches are defined as the activity between a stimulus and the relaxation to an inactive absorbing state.
On the other hand, experimental neuronal avalanches are defined by the activity between consecutive silent states.
We claim that the latter definition may be extended to some theoretical models in order to characterize their power-law avalanches and critical behavior.
We study a system in which the separation of driving and relaxation time scales emerges from its structure.
We apply both definitions of avalanche to our model. Both yield power-law distributed avalanches that scale with system size
in the critical point as expected.
Nevertheless, we find restricted power-law distributed avalanches outside of the critical region within the experimental procedure,
which is not expected by the standard theoretical definition. We remark that these results are dependent on the model details.
\end{abstract}

\pacs{05.70.Jk,45.70.Ht,87.18.Hf,87.19.lj}

\maketitle

\section{Introduction}

Power laws (PL) are ubiquitous in nature~\cite{bakSOC}, appearing in the description of phenomena as diverse as
sand pile avalanches~\cite{jensenSOC,pruessnerSOC2012}, rice pile avalanches~\cite{fretteArroz},
forest fires~\cite{christensenFFM,malamudFFM}, the velocity distribution of ants in ant farms~\cite{christensenAnts},
the magnitude distribution of earthquakes~\cite{gutenbergRichter1949},
surface growth~\cite{jensenSOC}, superconductor vortices~\cite{jensenSOC}, solar flares~\cite{wangSOCAstro2013},
stock market prices~\cite{gabaixPLecono2003}, connectivity of social networks~\cite{barabasiReview},
evolution, ecosystems, epidemics~\cite{gisiger2001},
neuronal avalanches~\cite{chialvoReview,plenzBenefits,beggsCritical}, etc.
Critical phenomena are characterized by PL in two different ways.
In the critical point, the correlation function decays as a PL.
Close to a critical point, the order parameter, susceptibility
and other thermodynamical quantities have PL behavior, governed
by critical exponents~\cite{stanley}.
Inspired by the theory of critical phenomena~\cite{stanley}, \autorano{bakPRL} proposed that these phenomena's events are PL
distributed because such systems self-organize to a non-equilibrium critical point. This was the inception of the
theory of self-organized criticality (SOC)~\cite{jensenSOC,pruessnerSOC2012}.

The authors used a sand pile-like cellular automaton model to describe the dynamics of such systems: at each time step a grain of sand
is added to a random pile inside a square lattice. When a pile reaches a predefined critical slope, it redistributes four of its exceeding
grains of sand to its four neighbors; if any of the neighbors becomes unstable, it also redistributes its grains and so on, interactively,
until the system reaches a new absorbing configuration.
The result of this branching process is an avalanche of size $\sabs$ (total number of redistributed grains)
and duration of $\Tabs$ time steps. As long as there is an
avalanche happening, the external deposition of grains into the system is paused. It restarts only after the avalanche has ceased.
This procedure is known as the separation of time scales of external driving and internal relaxation.
Avalanches thus generated have sizes and duration distributed according to PL:
$\Pdist(\sabs)\sim\sabs^{-\aabs}$ and $\Pdist(\Tabs)\sim\Tabs^{-\tabs}$, where size and duration are expected
to satisfy $\sabs\sim\Tabs^{a}$ such that $\aabs$ and $\tabs$ follow Sethna's
scaling relation $a=(\tabs-1)/(\aabs-1)$~\cite{sethna2001}.

In the critical point, even in the presence of short range interactions, an initial stimulus propagates throughout
the system because of the PL long-range correlation in space and time.
It produces a $1/f$ power spectrum signature in many of these systems~\cite{jensenSOC}. The external separation of time
scales breaks down the temporal correlations between subsequent avalanches, so it is not surprising that the sand pile model generates
$1/f^2$ power spectrum instead~\cite{jensen1f}. The systems which have sand pile-like dynamics are known to pertain to
universality classes of phase transitions into absorbing states~\cite{pruessnerSOC2012}.

SOC has been proposed to describe the brain activity~\cite{stassiBrain,usher1f1995} and the
first PL distributed neuronal avalanches were found by \autorano{beggsPlenz2003} using \textit{in vitro} local field potentials (LFP)
recordings. Many models were proposed to generate avalanches using brain-inspired
systems~\cite{kinouchiCopelli,socPlasticity,levina,ariadneDynSyn2015,ribeiroCopelli,girardiAva,beggsSOqC2014,shewV1Aval2015}.
Usually, models rely on probabilistic cellular automata, dynamical maps, or integrate-and-fire neurons connected through dynamical or
stochastic synapses.
All of these models, however, are similar to the sand pile model in the sense that they independently generate avalanches,
forcefully separating stimulation and relaxation time scales.
Consequently, the long-range correlations~\cite{linkenkaerSOC,haimoviciRestState2013} and
$1/f$ power spectrum~\cite{novikov1fB1997,teichcat97,linkenkaerSOC,henrie1fV12005,hermes1fV12014} found experimentally for the
healthy state of the brain are not well described by most of these models. Notice, however, that some models on these classes
may exhibit $1/f$-like power spectrum close to the critical state by the introduction of inhibitory synapses~\cite{socPlasticity,lombardi1f2017}.

The experimental procedure for measuring neuronal avalanches consists of low or high-pass filtering each LFP time series, and then
thresholding them in order to obtain a time series of spike-like activity.
The data are then binned in time using the average interevent interval (each event is a spike-like activity in any electrode
processed signal). Empty time bins are used to separate subsequent avalanches. Notice that these empty
time bins are not analogous to the inactive absorbing state used to separate avalanches in each of the neuronal avalanches models
listed above due to the ignored subthreshold background activity of the electrodes.
Such method has been applied to many works probing for neuronal
avalanches~\cite{beggsPlenz2004,violaSub,ribeiroCopelli,plenzAvalV12010,priesemannSub2014,shewV1Aval2015}.
In fact, experimental evidence suggests that a better analogous to such LFP recordings would be a sand pile subjected to
random background activity due to almost absence of empty time bins~\cite{priesemannSub2014}.

A reasonable extension of the experimental approach can be applied to theoretical models in order to probe for power-law
distributed avalanches and the critical state~\cite{scarpettaMem2013,scarpettaMem2014,hidalgo2017}.
However, the two approaches may yield different results~\cite{hidalgo2017}. In this work, we present a model in which
both approaches agree in the determination of the critical state. Moreover, the second approach allows the calculation of
the power spectrum for our particular model and seems to yield restricted power-law avalanches even outside of the critical range.
Notice that these results are dependent of model parameters.

Here we study a biologically plausible mammalian visual system model~\cite{girardiV12015}. It presents an absorbing phase transition to a
percolating phase in which the activity gets spontaneously separated in time due to its intrinsic dynamics.
An ambiguity concerning the definition of avalanches
is naturally raised: should avalanche be defined as the activity between the stimulus and the moment the system reaches the
inactive absorbing state or as the network activity between consecutive silent states?
We compare both approaches and show that inside the critical range they are very similar and both
display PL distributed avalanches with diverging cutoff. However, outside the critical range the second approach still yields restricted PL
distributed avalanches whereas the first approach yields the expected break down of PL behavior. In fact,
we show that inside the critical range, the power spectrum of the first approach is white-noise-like, while the second
approach presents approximately $1/f$ noise~\cite{girardiV12015}.

\section{Model}

\begin{figure}[t!]
    \centerline{
	\includegraphics[width=0.4\textwidth]{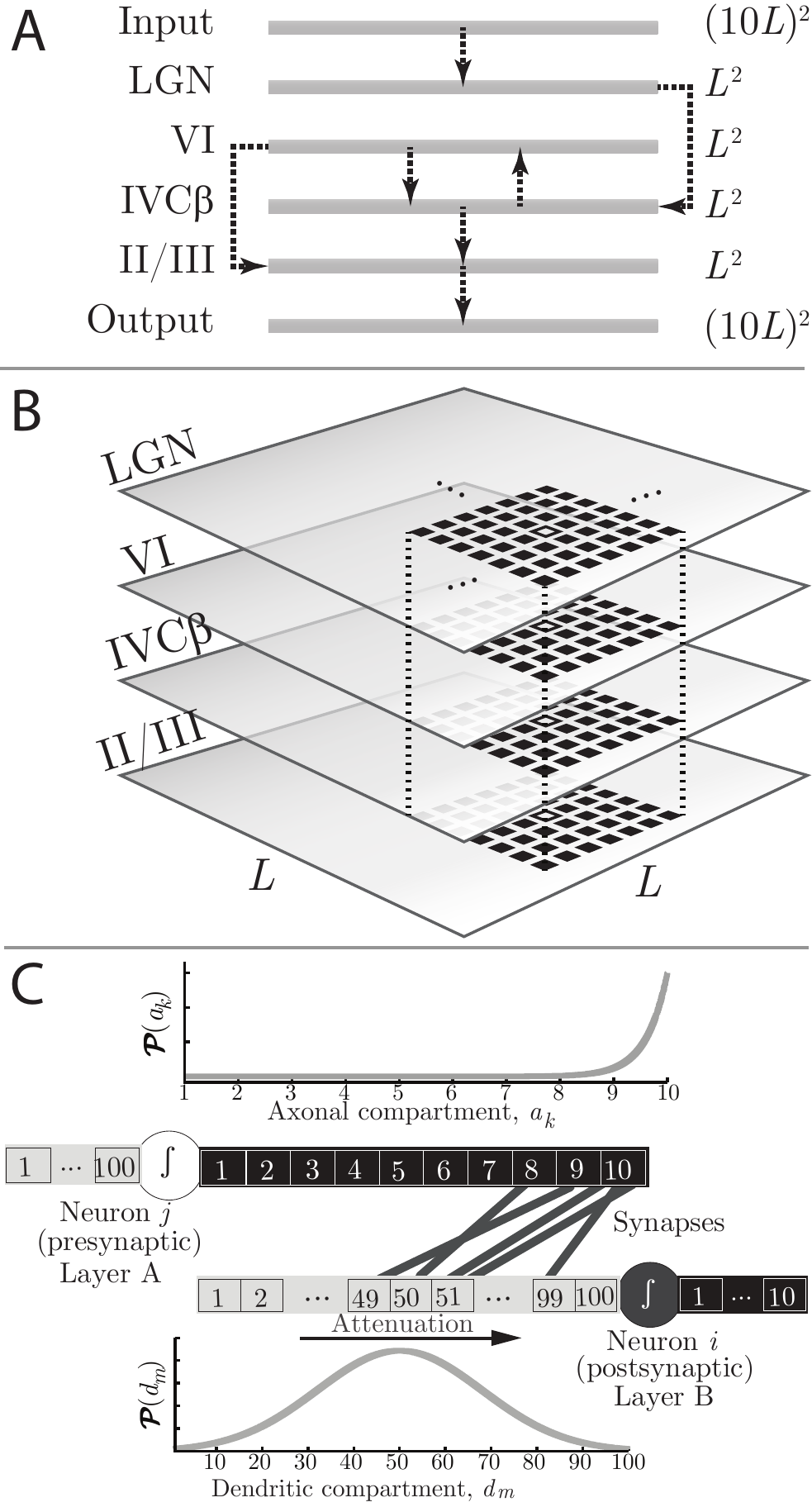}
    }
	\caption{\label{fig:modeloV1}Elements of the V1 model.
	A: Architecture of the network.
    B: Spatial organization of the network of $N=4L^2$ neurons. The columnar structure is highlighted in between dotted lines.
    There is a column of size $N_c=4l^2=196$ neurons centered on each neuron of the network.
    C: The compartmental structure of the neuron: dendritic compartments (light gray), axonal compartments (black) and soma (dark gray).
    The synapses are formed between presynaptic axonal compartments [chosen according to an exponential probability
    distribution $\mathcal{P}(a_k)$] and dendritic compartments [chosen with Gaussian probability distribution $\mathcal{P}(d_m)$].
    The signal attenuates while moving forward in the dendrites.}
\end{figure}

The model has two components: structure and dynamics. The structure is based on the form recognition pathway of the
mammalian visual system~\cite{girardiV12015}. There are six square layers sequentially stacked one on top of the other
(Fig.~\ref{fig:modeloV1}A and~B). The top layer is the input where each of its elements represents a photoreceptor of the retina.
The bottom layer is the output of the
primary visual cortex (V1) where each element represents one axonal terminal that would connect to secondary visual area of the cortex.
The input and output layers have $N_{io}=(10L)^2$ elements each. The four internal layers are the \camLGN\ from the thalamus,
and the \camVI, \camIV\ and \camII@III\ from V1, respectively. Each of these four layers contains $L^2$ neurons. The network
thus has $N=4L^2$ neurons. Each neuron is composed of 100 dendritic compartments, 1 soma compartment and 10 axonal compartments
(Fig.~\ref{fig:modeloV1}C).

The network is built according to the following algorithm: for each neuron of the network,
(a) a postsynaptic neuron is drawn from an adjacent layer (according to arrows in Fig.~\ref{fig:modeloV1}A) inside a limited
excitatory field of $7\times7$ neurons (columns in Fig.~\ref{fig:modeloV1}B); a bidimensional Gaussian probability distribution centered in
the presynaptic neuron with standard deviation $3$ is used to draw the adjacent postsynaptic neuron;
(b) an axonal compartment is drawn from the presynaptic neuron using an exponential probability distribution
(the farther from the soma the more probable it is to be chosen -- Fig.~\ref{fig:modeloV1}C);
(c) a dendritic compartment is drawn from the postsynaptic neuron using a
Gaussian probability distribution centered in the center of the dendrite with standard deviation of $20$ dendritic
compartments (Fig.~\ref{fig:modeloV1}C);
(d) the chosen axonal compartment is then connected to the chosen dendritic compartment. Each element of the input layer
is connected to a dendritic compartment a neuron of the \camLGN\ in the same way that a presynaptic axonal compartment would be.
The amount of synapses for the adjacent layers are presented in Table~\ref{tab:synPerPreSyn}.
Notice that the limited excitatory field of presynaptic neurons in step (a) generates the well known columnar structure in the
network~\cite{albrightNeuroRev2000}, illustrated in Fig.~\ref{fig:modeloV1}.
One trial of the simulation consists of: building the network, stimulating the network and waiting for the activity to cease.
This procedure gives rise to quenched disorder. Thus, many trials of the network disorder are necessary to calculate observables.

\begin{table}[b!]
\caption{\label{tab:synPerPreSyn}
Amount of connections of each layer of the model per presynaptic element. Connections come from layers listed as rows and go
to layers listed as columns of the table. Values are derived from many experimental works~\cite{girardiV12015}.}
\begin{center}
\begin{tabular}{c|ccccc}
\hline
\textbf{From~\textbackslash~To}    & \textbf{\camLGN} & \textbf{\camVI}  & \textbf{\camIV} & \textbf{\camII@III} & \textbf{Output}\\\hline\hline
\textbf{Input}      & 1 & -    & -    & -   & -   \\
\textbf{\camLGN}    & - & -    & 500  & -   & -   \\
\textbf{\camVI}     & - & -    & 1100 & 350 & -   \\
\textbf{\camIV}     & - & 600  & -    & 700 & -   \\
\textbf{\camII@III} & - & -    & -    & -   & 100 \\\hline
\end{tabular}
\end{center}
\end{table}

Initially, a $30\times30$ square of photoreceptors in the input layer is instantly activated, corresponding to approximately $3\times3$
neurons in the \camLGN (all the other neurons are quiescent). The initial stimulation is analogous to a flash presented to the retina.
The signal is then propagated one compartment per time step, from the dendrites to the last axonal compartment.
Each dendritic compartment sums up the excitatory postsynaptic potential (EPSP), $E$, from all its active presynaptic axonal compartments
with the signal from the previous dendritic compartment. This sum is attenuated at each time step by a constant $\lambda$.
A soma compartment fires an action potential if the signal in the last dendritic compartment is greater than a threshold $v_T$.
The axonal compartments propagate the signal coming from the soma without any dissipation. These rules are expressed as:

\begin{equation}
\label{eq:compDendritoV1}
d_k^{(i)}(t+1)=\lambda\left[ d_{k-1}^{(i)}(t)+E\sum\limits_{j,n}{a_n^{(j)}(t)}\right]\ ,
\end{equation}
\begin{equation}
\label{eq:compSomaV1}
v_i(t+1)=\left\{
\begin{array}{ll}
    \Theta\left(d_{100}^{(i)}(t)-v_T\right) & \textnormal{, if }v_i(t)=0\virgula\\
    -V_R R & \textnormal{, if }v_i(t)=1\virgula\\
    v_i(t)+1 & \textnormal{, if }v_i(t)<0\virgula
\end{array}
\right.\ ,
\end{equation}
\begin{equation}
\label{eq:compAxonioV1}
a_k^{(i)}(t+1)=\Theta\left(v_i(t)\right)\delta_{k,1}+a_{k-1}^{(i)}(t)\ ,
\end{equation}
where $t$ is the time step, $d_k^{(i)}(t)$ [$a_k^{(i)}(t)$] is the potential in the dendritic [axonal] compartment $k$ of the neuron $i$
at time $t$, $v_i(t)$ is the soma potential of neuron $i$ at time $t$, $E>0$ is the EPSP elicited by each axonal compartment $n$ of
presynaptic neuron $j$, $\lambda=0.996$ is the membrane attenuation constant, $v_T=10\mV$ is the firing threshold, $R$ is the refractory period, $V_R\equiv1$ is
a dimensionality constant measured in mili-volts per time steps,
$\Theta(x)$ is the Heaviside step function and $\delta_{m,n}$ is the Kronecker delta. Notice that $d_0(t)=a_0(t)\equiv0$ for all the
neurons.

The balance between excitation (controlled by $E$), dissipation (controlled by $\lambda$) and the refractory period $R$ determines
whether the system is in an inactive, active or percolating phase~\cite{girardiV12015}. $R$ is adjusted such that there is no
self-sustained activity in the interlayer loop between layers \camVI\ and \camIV. $\lambda$ is fitted to the experimental value
obtained in Ref.~\cite{williamsV12002}. Under these conditions, $E$ controls a phase transition between an inactive absorbing state
(subcritical activity, $E<1.11\mV$) and an inactive percolating phase (supercritical activity, $E>1.19\mV$).
The range $1.11\leq E\leq1.19\mV$ is an extended region of critical behavior~\cite{girardiV12015}.

\begin{figure}[t!]
    \centerline{
	\includegraphics[width=0.45\textwidth]{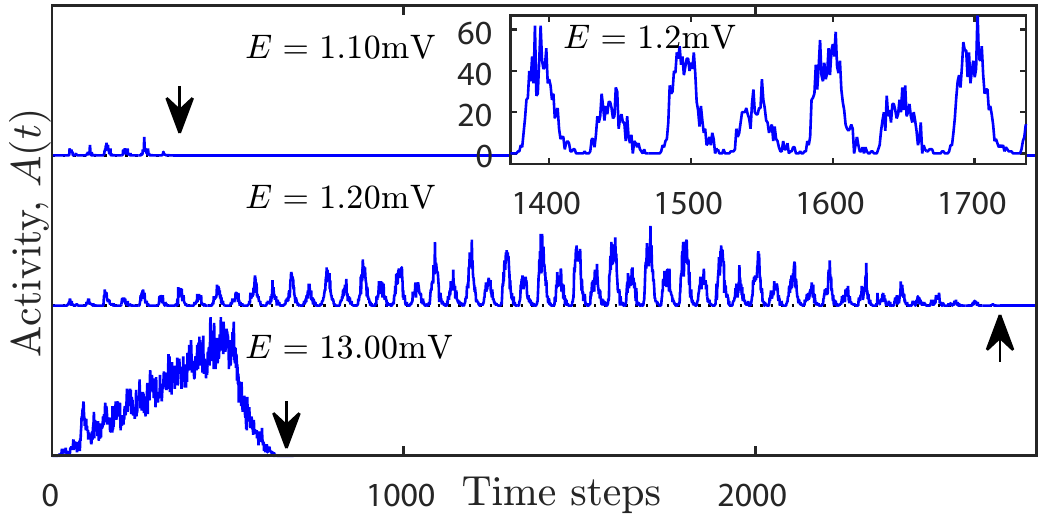}
    }
	\caption{\label{fig:activity}Sum of all network spikes, $A(t)$, for each time step $t$ and three typical values of
    $E$. Inset: detail of the network activity for $E=1.2\mV$ highlighting the silent periods between peaks of activity.
    Arrows mark the end of the activity generated by a single stimulus.}
\end{figure}

\section{Results}

The network activity for a given simulation trial, $A(t)$, is simply the sum of every soma spike at each time step $t$,
\begin{equation}
\label{eq:activity}
A(t)=\sum_{i=1}^{N}\delta_{v_i(t),1}\ ,
\end{equation}
where $\delta_{v_i(t),1}=1$ if $v_i(t)=1$ (zero otherwise) is the Kronecker delta.
The profile of $A(t)$ in Fig.~\ref{fig:activity} allows us to define avalanches in two different ways:
(I) an avalanche is regarded as all the activity due to a single stimulus, similarly to the procedure applied to
the sand pile model and other absorbing state systems~\cite{bakPRL,pruessnerSOC2012}. In this approach,
each curve in Fig.~\ref{fig:activity} for a typical $E$ represents a single avalanche. In order to generate other avalanches,
we must perform other simulation trials (i.e. rebuilding the network and re-stimulating the Input layer).
(II) an avalanche is all the activity surrounded by silent periods, similarly to the experimental procedure used to measure
neuronal avalanches. Within this second approach, each curve in Fig.~\ref{fig:activity} for a typical $E$ contains many avalanches
because each avalanche corresponds to an activity peak (which are highlighted in the inset of this figure for $E=1.2\mV$).

We select three values of EPSP (subcritical, $E=1.1\mV$; critical, $E=1.185\mV$; and supercritical $E=13\mV$)
in order to study the shape of avalanche distributions, $\Pdist$, and complementary
cumulative distributions, $\Fdist$, of both avalanche sizes, $s$, and duration, $T$.
The avalanche sizes and duration distributions are expected to scale according to~\cite{pruessnerSOC2012}:
\begin{align}
\label{eq:distS}
\Pdist(s)&\sim s^{-\alpha}\Gsc_s(s/s_c)\ ,\\
\label{eq:distT}
\Pdist(T)&\sim T^{-\tau}\Gsc_T(T/T_c)\ ,
\end{align}
where $\alpha$ and $\tau$ are scaling exponents, $\Gsc_{s,T}(x)$ are scaling functions and $s_c$ and $T_c$ are the cutoff of the
distributions. In the critical point, both cutoff are expected to scale with system size:
\begin{align}
\label{eq:cutoffS}
s_c\sim L^D\ ,\\
\label{eq:cutoffT}
T_c\sim L^{\mu}\ ,
\end{align}
where $L$ is the characteristic system size, $D$ ($\mu$) is a characteristic dimension of the avalanche sizes
(duration)~\cite{pruessnerSOC2012}.
In fact, the PL behavior holds only for $s<s_c$ (or $T<T_c$) in Eqs.~\eqref{eq:distS} and~\eqref{eq:distT}.

\begin{figure}[t!]
    \centerline{
	\includegraphics[width=0.45\textwidth]{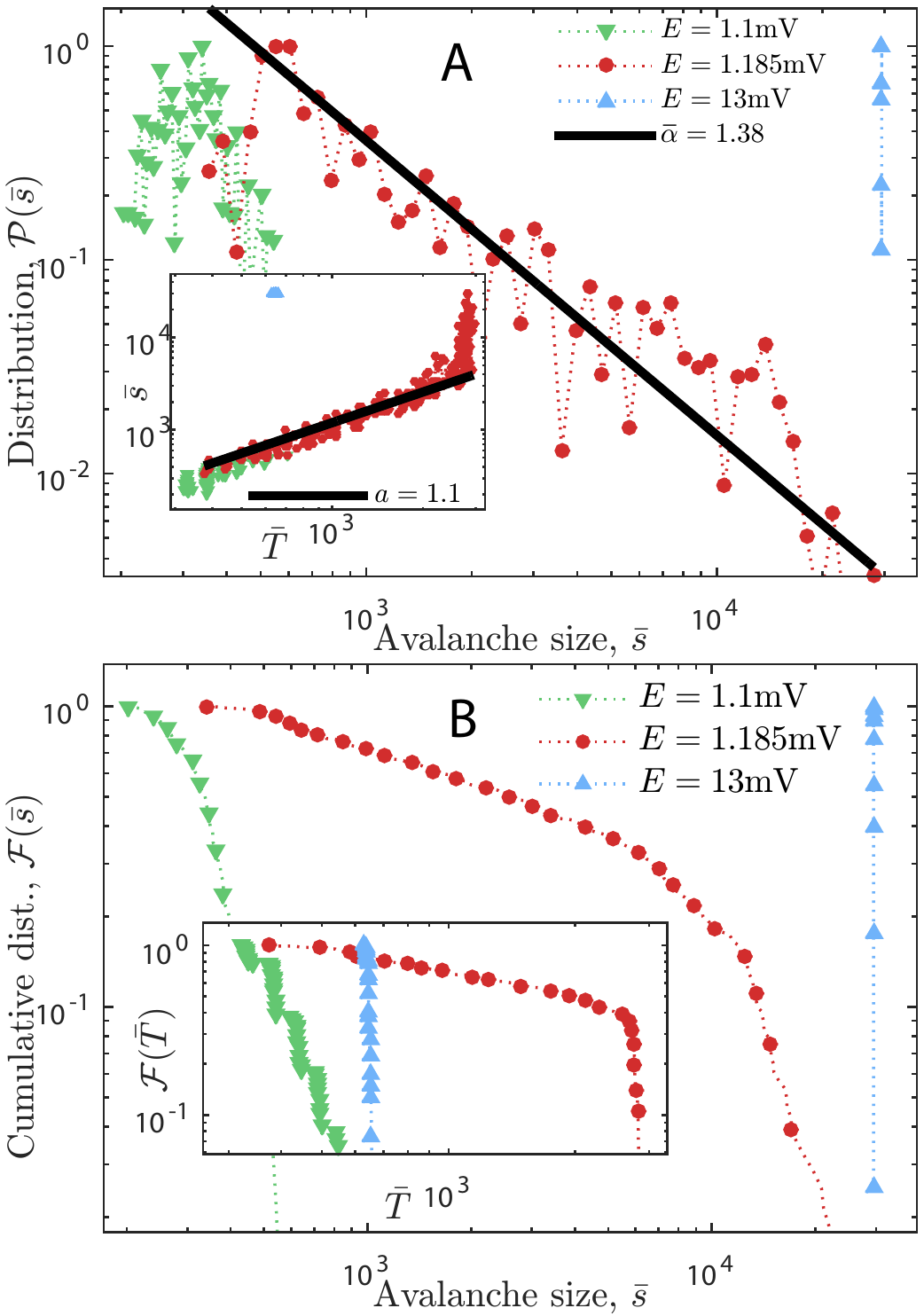}
    }
	\caption{\label{fig:avAbsPs}Distributions for the absorbing state avalanches (approach I)
    in three $E$ regimes: (\textcolor{greenC}{$\blacktriangledown$}) subcritical, (\textcolor{redC}{\ding{108}}) critical, and
    (\textcolor{blueC}{$\blacktriangle$}) supercritical.
	A: $\Pdist(\sabs)$ and PL fit for the critical regime yielding $\aabs=1.38$. A~inset: Sethna's scaling relation $\sabs\sim\Tabs^a$
    yields $a=1.1(1)$ for the critical regime.
    B: $\Fdist(\sabs)$ and $\Fdist(\Tabs)$ for the three considered regimes. Notice that only the critical regime has PL distributed
    avalanches.}
\end{figure}

The complementary cumulative distribution (henceforth referred only as cumulative distribution) is defined as
$\Fdist(s)=\int_{s}^{\infty}\Pdist(s')ds'$. In this work we rather choose to analyze the cumulative distribution because
it provides a clearer visualization of the data, since it is a continuous function of
its variables. It also displays reduced noise because its precision does not depend on the size of the bins of the distribution’s 
histogram and it has a better defined cutoff~\cite{newmanPowerlaw}. 
We may write $\Fdist(s)$ and $\Fdist(T)$ using scaling functions,
\begin{align}
\label{eq:cdistS}
\Fdist(s)&\sim s^{-\alpha+1}\Hsc_s(s/s_c)\ ,\\
\label{eq:cdistT}
\Fdist(T)&\sim T^{-\tau+1}\Hsc_T(T/T_c)\ ,
\end{align}
where $\Hsc_{s,T}(x)$ are scaling functions and $s_c$ and $T_c$ are defined in Eqs.~\eqref{eq:cutoffS} and~\eqref{eq:cutoffT}.

The exponents $\alpha$, $\tau$, $D$ and $\mu$ may be calculated through the collapse of the distributions~\cite{pruessnerSOC2012}:
we run trials for many different system sizes, $L$, and then plot
$s^{\alpha-1}\Fdist(s)$ versus $s/L^D$; the best value of $\alpha$ collapses the data due to different $L$ over a single horizontal line,
whilst the best value of $D$ collapses the cutoff of the distributions due to different $L$ over a single vertical line.
This procedure may also be applied to $\Fdist(T)$.

Another way to calculate these exponents is to use the fact that the PL holds only for $s<s_c$ and write $\Fdist(s)$
as~\cite{girardiAva}
\begin{equation}
\label{eq:cDistFit}
\Fdist(s)=\int_{s}^{s_c}{\Pdist(s')ds'}=c_1+c_2s^{-\alpha+1}\ ,
\end{equation}
where the cutoff may be directly calculated by fitting $c_1$, $c_2$ and $\alpha$ to the computational data using Eq.~\eqref{eq:cDistFit}:
$s_c=(-c_1/c_2)^{1/(-\alpha+1)}$. The same equation may be fitted for $\Fdist(T)$ yielding $\tau$ and $T_c$.
We will use a bar over $s$, $T$, $\alpha$ and $\tau$ to denote measurements using approach I.

\subsection{Absorbing state avalanches}

\begin{figure}[b!]
    \centerline{
	\includegraphics[width=0.45\textwidth]{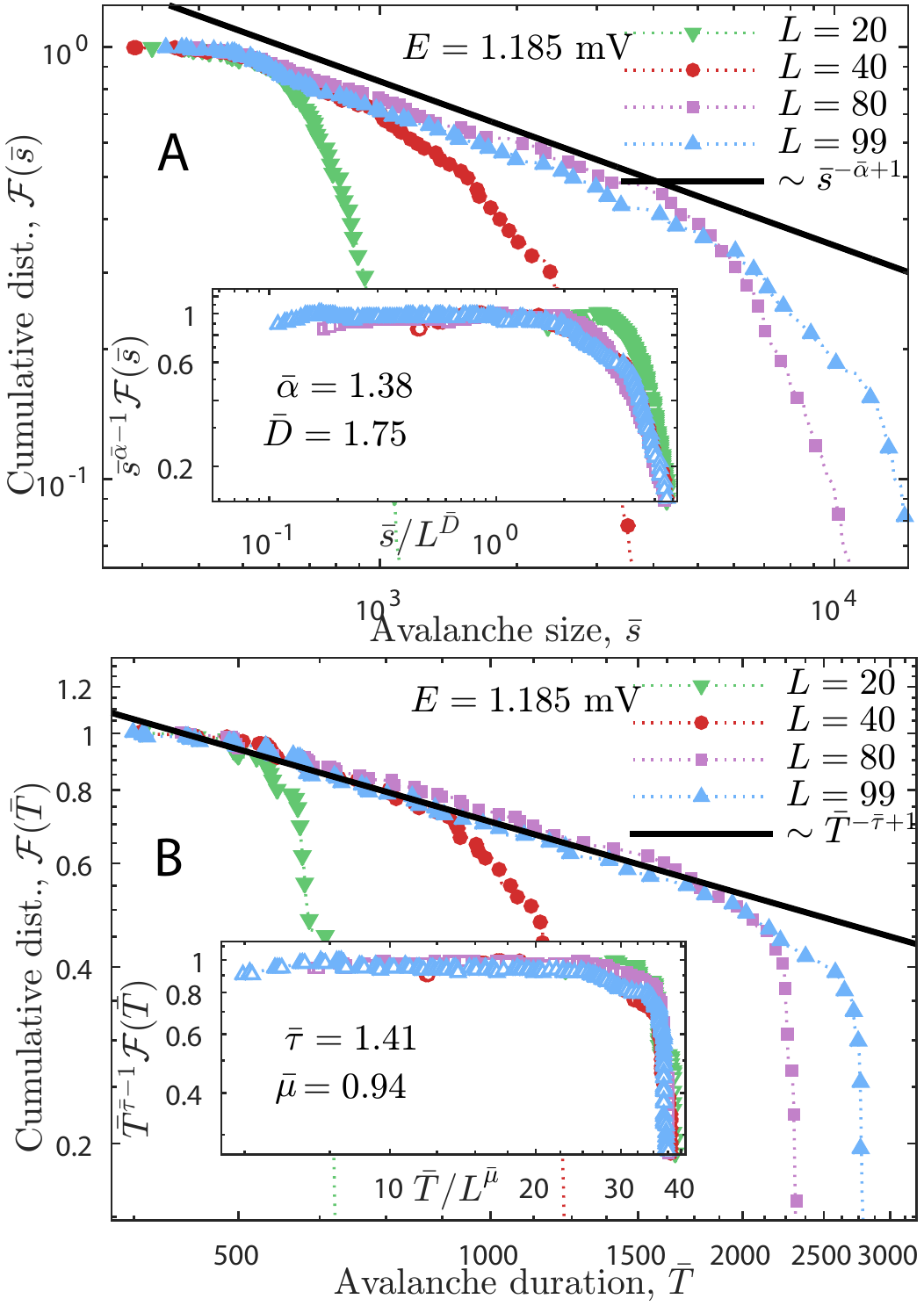}
    }
	\caption{\label{fig:avAbsFsT}Cumulative distributions of absorbing state avalanches (approach I) for the critical regime ($E=1.185\mV$)
    and four network sizes: (\textcolor{greenC}{$\blacktriangledown$}) $L=20$, (\textcolor{redC}{\ding{108}}) $L=40$,
    (\textcolor{purpleC}{$\blacksquare$}) $L=80$ and (\textcolor{blueC}{$\blacktriangle$}) $L=99$.
	A: $\Fdist(\sabs)$. A~inset: plot of $\Hsc(\sabs/L^{\Dabs})\sim\sabs^{\aabs-1}\Fdist(\sabs)$ (collapse of the distributions) yielding
    $\aabs=1.38$ and $\Dabs=1.75$.
    B: $\Fdist(\Tabs)$. B~inset: plot of $\Hsc(\Tabs/L^{\muabs})\sim\Tabs^{\tabs-1}\Fdist(\Tabs)$ (collapse of the distributions) yielding
    $\tabs=1.41$ and $\muabs=0.94$.}
\end{figure}

We define the absorbing state (approach I) avalanche size, $\sabs$, as the sum of all the spikes generated by a single stimulus
(i.e. by a single simulation trial):
\begin{equation}
\label{eq:sAbs}
\sabs=\sum_{t=0}^{\Tabs}A(t)\ ,
\end{equation}
where $\Tabs$ is the avalanche duration marked by arrows in Fig.~\ref{fig:activity}.
$\Tabs$ is the time needed for the system to relax into an absorbing inactive state. Many trials of the network are considered,
each one yields an avalanche with size $\sabs$ and duration $\Tabs$.
Fig.~\ref{fig:avAbsPs}A (Fig.~\ref{fig:avAbsPs}B) shows the distribution (cumulative distribution) of avalanche sizes
(sizes and duration) for three different
$E$: upside-down green triangles correspond to the subcritical phase, red circles correspond to the critical phase and blue triangles
correspond to the supercritical phase.

The PL distributions, $\Pdist(\sabs)\sim\sabs^{-\aabs}$ and $\Pdist(\Tabs)\sim\Tabs^{-\tabs}$, from Eqs.~\eqref{eq:distS}
and~\eqref{eq:distT}, are obtained only inside the critical range $1.11\leq E\leq1.19\mV$ as expected.
The inset of Fig.~\ref{fig:avAbsPs}A shows that the scaling relation $\sabs\sim\Tabs^a$ with $a=1.1(2)$ holds for the critical system.
Such relation is derived assuming that both $\Pdist(s)$ and $\Pdist(T)$ are PL shaped.
Fig.~\ref{fig:avAbsFsT}A and B show the cumulative distributions of avalanche size and duration for many system sizes
in the critical range for $E=1.185\mV$. The collapse of the cumulative distributions yields $\aabs=1.38$ and $\Dabs=1.75$ for
the avalanche sizes, and $\tabs=1.41$ and $\muabs=0.94$ for the avalanche duration [Eqs.~\eqref{eq:cutoffS} to~\eqref{eq:cdistT}].
Sethna's scaling relation~\cite{sethna2001}, $a=(\tabs-1)/(\aabs-1)=0.41/0.38\approx1.07$, agrees very well with the fitted value in
Fig.~\ref{fig:avAbsPs}A inset.

\begin{figure}[t!]
    \centerline{
	\includegraphics[width=0.45\textwidth]{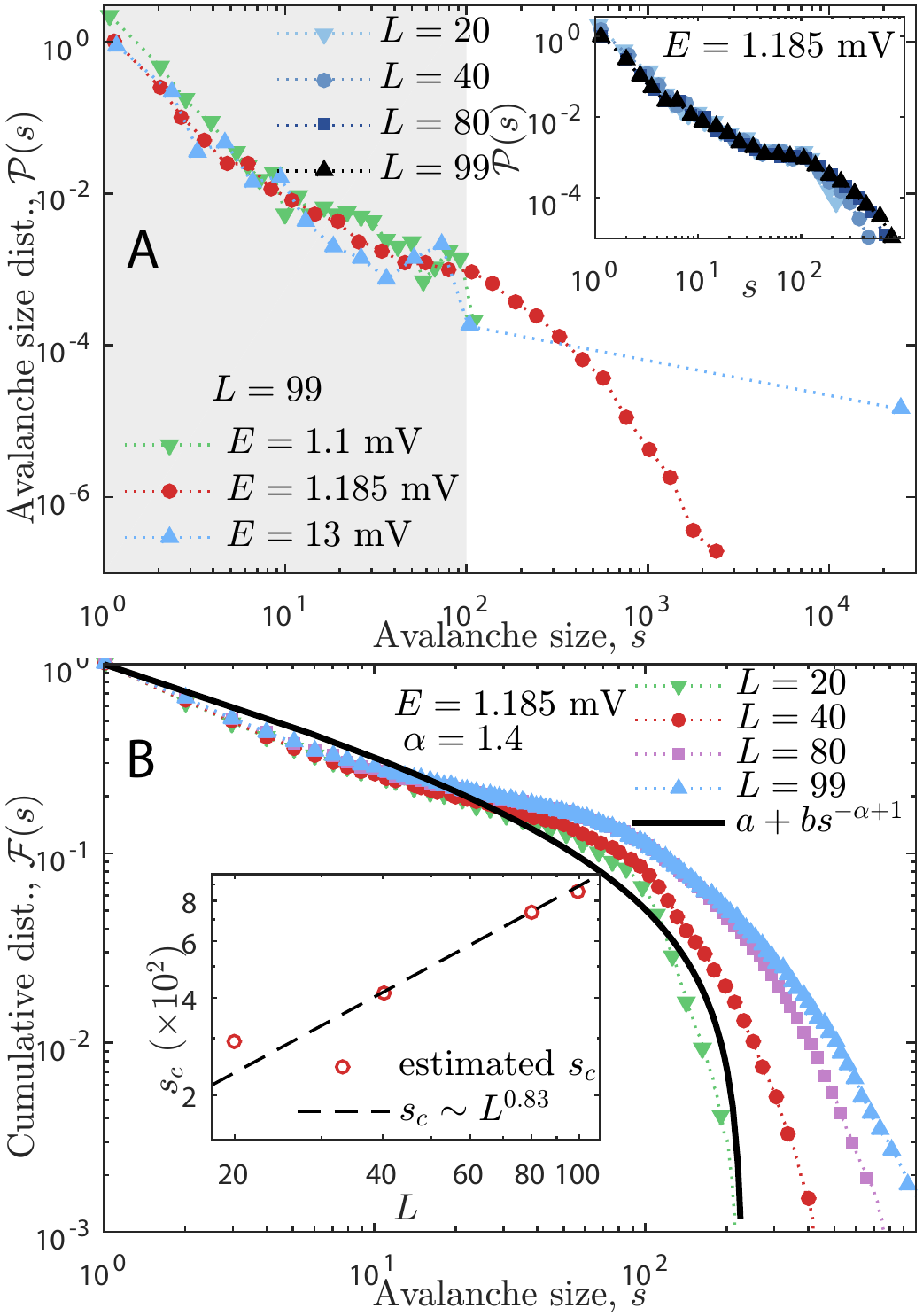}
    }
	\caption{\label{fig:avSilS}Silent periods avalanche size distributions (approach II).
	A: $\Pdist(s)$ for $E$ in three regimes: (\textcolor{greenC}{$\blacktriangledown$}) subcritical,
    (\textcolor{redC}{\ding{108}}) critical, and (\textcolor{blueC}{$\blacktriangle$}) supercritical. Notice that the three regimes
    have PL shape inside the shaded area.
    A~inset: critical regime $\Pdist(s)$ for critical $E=1.185\mV$ and different $L$.
    B: $\Fdist(s)$ for critical $E=1.185\mV$ and different $L$ used to fit Eq.~\eqref{eq:cDistFit} yielding
    $s_c(L)$ and $\alpha=1.4$. Solid line: example fit of Eq.~\eqref{eq:cDistFit} for $L=20$ (R-Squared: $99.9\%$).
    B~inset: cutoff $s_c\sim L^D$ yielding $D=0.8(2)$.}
\end{figure}

\subsection{Silent periods avalanches}

The silent periods (approach II) avalanche size, $s$, is defined as the sum of all the spikes of the network between two consecutive
periods of silence:
\begin{equation}
s=\sum_{t_n}^{t_{n+1}}A(t)\ ,
\end{equation}
where the $t_n$ are such that $A(t_n)=A(t_{n+1})=0$ and $A(t_n<t<t_{n+1})>0$. The avalanche duration is
simply the number of time steps in between two consecutive moments of silence, $T=t_{n+1}-t_n$~\cite{girardiV1conf2015,girardiV12015}.

\begin{figure}[b!]
    \centerline{
	\includegraphics[width=0.45\textwidth]{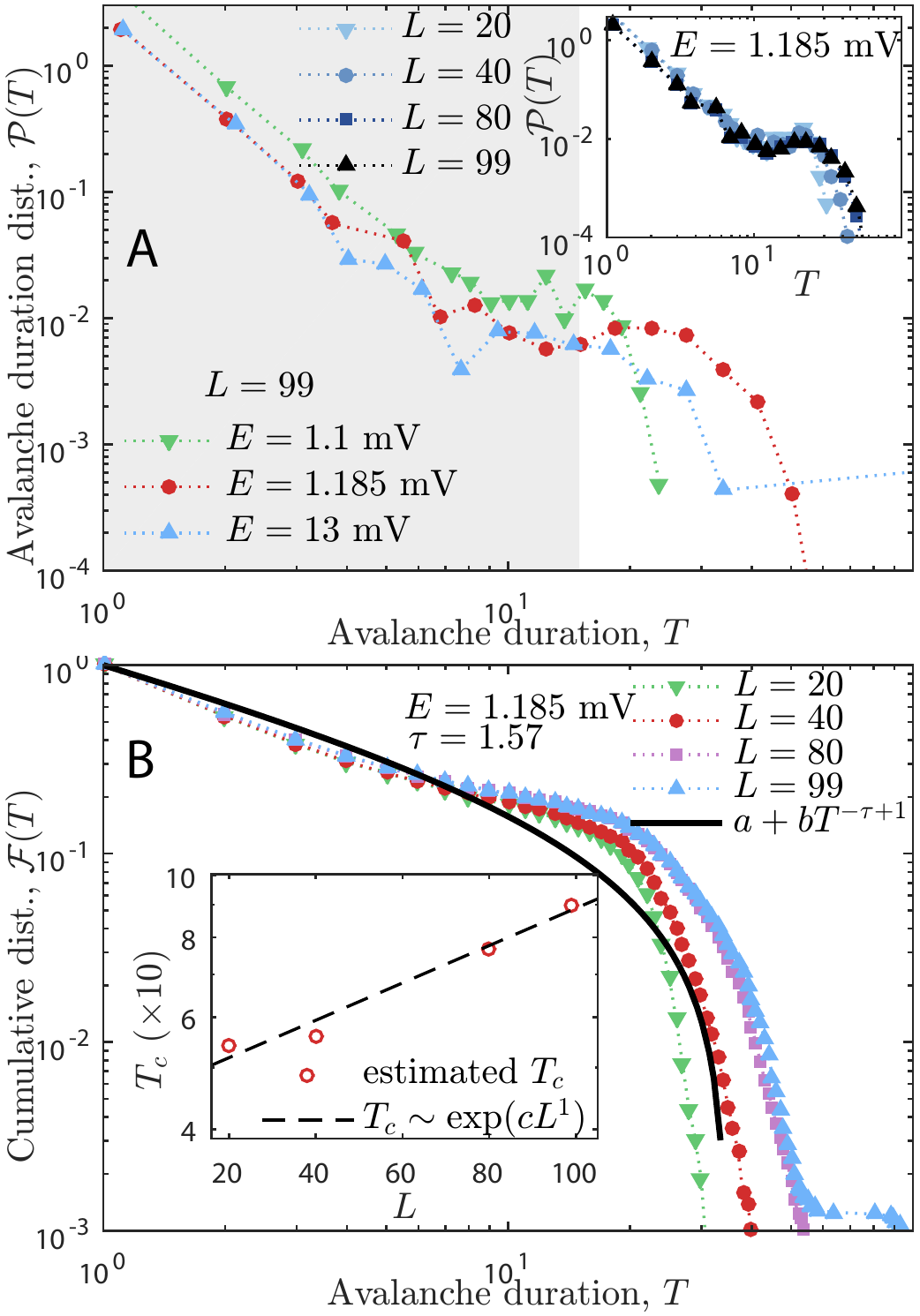}
    }
	\caption{\label{fig:avSilT}Silent periods avalanche duration distributions (approach II).
	A: $\Pdist(T)$ for $E$ in three regimes: (\textcolor{greenC}{$\blacktriangledown$}) subcritical,
    (\textcolor{redC}{\ding{108}}) critical, and (\textcolor{blueC}{$\blacktriangle$}) supercritical. Notice that the three regimes
    have PL shape inside the shaded area.
    A~inset: critical regime $\Pdist(T)$ for critical $E=1.185\mV$ and different $L$.
    B: $\Fdist(T)$ for critical $E=1.185\mV$ and different $L$ used to fit Eq.~\eqref{eq:cDistFit} yielding
    $T_c(L)$ and $\tau=1.57$. Solid line: example fit of Eq.~\eqref{eq:cDistFit} for $L=20$ (R-Squared: $99.4\%$).
    B~inset: cutoff $T_c\sim\exp(cL^D)$ yielding $D=1.0(2)$.}
\end{figure}

The interspersing peaks of activity emerge in this model due to the deterministic dynamics and compartmental body of the neurons.
The apparent inactive moments happen while the signal is being propagated through dendrites and axons. The duration of such
silent periods depends on network architecture and neurons parameters, such as the amount of dendritic and axonal compartments and
the balance between the refractory period, $R$, and the excitation-dissipation rate (controlled by $E$ and $\lambda$).
The separation of avalanches is then very natural because there is no need to reset the system in order to spark another
wave of activity. Similarly, the ignored background activity in experimental measurements (and hence the experimental separation
of time scales) contains weak electric signals generated by the propagation of potentials through membranes and synapses.

In this work, we are interested in comparing the avalanches thus defined with those of approach I.
Notice that the avalanche sizes and duration under approach II have been thoroughly studied and shown to be PL distributed in the critical region of this
model (see Figs.~\ref{fig:avSilS} and~\ref{fig:avSilT})~\cite{girardiV12015,girardiV1conf2015}.
The cutoff of such distributions, calculated here using the fit of Eq.~\eqref{eq:cDistFit} to different values of $L$, also non-trivially scale with system size,
as expected for regular absorbing state avalanches (inset of Figs.~\ref{fig:avSilS}B and~\ref{fig:avSilT}B)~\cite{girardiV12015}.
However, small avalanches are also PL distributed for sub and supercritical values of $E$
(see the shaded area of Figs.~\ref{fig:avSilS}A and~\ref{fig:avSilT}A)~\cite{girardiV12015}.
The distributions also obey Sethna's scaling relation for all regimes of $E$~\cite{girardiV1conf2015}.
This result contrasts with the distributions obtained for absorbing state avalanches, in which only critical state avalanches
display PL shape, non-trivial finite size scaling, and obey Sethna's relation.

The scaling exponent $D$ of the avalanche size cutoff, $s_c$ (Fig.~\ref{fig:avSilS}B inset), reveals the dimensionality of the path that propagates
the avalanches~\cite{girardiV12015}: $D=0$ in the subcritical regime means that only small avalanches that remain confined in the bulk
of the network (independently of $L$) exist; $D\approx1$ in the critical regime means that avalanches are spreading preferentially inside
the columns of the network; $D\approx3$ and $D=2$ in the supercritical regime means that avalanches are either spreading both
radially and inside the columns or only radially as a surface wave~\cite{girardiV1conf2015,girardiV12015}.

Figs.~\ref{fig:avSilS}A and~\ref{fig:avSilT}A show that there is indeed a PL shape for small avalanche sizes and duration for every
considered $E$ (either subcritical, critical or supercritical). The inset of these figures show that the cutoff of these distributions
scale with system size $L$ for critical $E=1.185\mV$. Figs.~\ref{fig:avSilS}B and~\ref{fig:avSilT}B show the cumulative distributions for
the critical $E=1.185\mV$ and many $L$ (the same cases as the insets of panels A in both figures), in which it is easier to
notice the cutoff diverging as the system is increased. We fit Eq.~\eqref{eq:cDistFit} to every cumulative distribution
in Figs.~\ref{fig:avSilS}B and~\ref{fig:avSilT}B to estimate the cutoffs $s_c$ and $T_c$. These cutoffs are plotted against $L$
in the inset of both panels B of the respective figures. $s_c$ presents the usual behavior described by Eq.~\eqref{eq:cutoffS}
with $D=0.9(1)$ whereas $T_c$ presents a stretched exponential scaling with $L$, $T_c\sim\exp(kL^D)$ ($k$ is a constant)
with $D=1.0(1)$, matching the autocorrelation characteristic time of the system inside the Griffiths
phase~\cite{brayGrif1988,girardiV12015}. In fact, this system has been shown to present a Griffiths region for
the range $1.11\leq E\leq1.19(1)\mV$~\cite{girardiV12015}.

\subsection{Power spectra}

\begin{figure}[b!]
    \centerline{
	\includegraphics[width=0.45\textwidth]{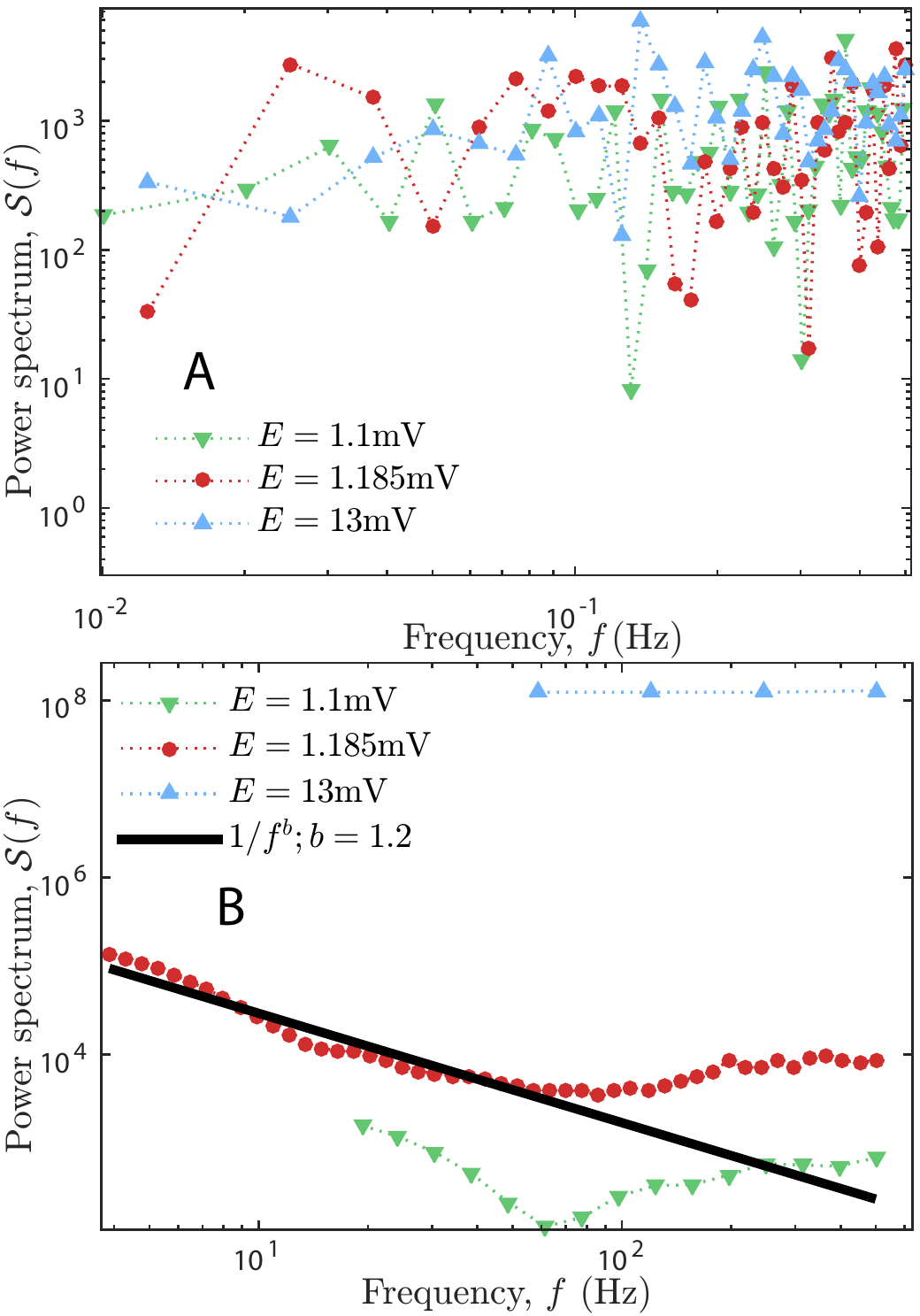}
    }
	\caption{\label{fig:avPS}Power spectra of avalanche size time series for both approaches and
    three regimes of $E$:
    (\textcolor{greenC}{$\blacktriangledown$}) subcritical, (\textcolor{redC}{\ding{108}}) critical, and
    (\textcolor{blueC}{$\blacktriangle$}) supercritical.
	A: $\Sps(f)$ for absorbing state avalanches (approach I). The three regimes have white noise power spectrum.
    B: $\Sps(f)$ for silent periods avalanches (approach II). The critical regime has $\Sps\sim1/f^b$ with $b\sim1.2$.}
\end{figure}

The power spectrum $\Sps(f)$ is related to temporal correlations via a Fourier transform. Thus, having a $\Sps\sim1/f$
means that there is long-range temporal correlations in the original signal. Such spectral density is found for the
healthy brain state using Magnetoencephalography and Electroencephalography
techniques~\cite{novikov1fB1997,teichcat97,linkenkaerSOC,henrie1fV12005,hermes1fV12014}.

The time series of avalanche sizes is constructed as follows: for approach I, the time series is generated by a sequence of stimuli,
such that each stimulus generates a single avalanche that spreads through a different realization of network disorder;
for approach II, the sequence of avalanche sizes is given by the sequence of the sums of the activity between two subsequent silent
intervals in Fig.~\ref{fig:activity}.
We calculated $\Sps(f)$ for the time series of avalanche sizes on the subcritical, critical and supercritical state.
The results are shown in Fig.~\ref{fig:avPS} for both approaches of measuring avalanches.

Absorbing state avalanches are generated by different stimuli on independent network disorder trials.
This approach is then expected to lack the $1/f$ profile because subsequent avalanches are statistically independent.
Indeed, Fig.~\ref{fig:avPS}A shows that all the considered $E$ have the exact same constant power spectrum, corresponding to
white noise.

On the other hand, silent periods avalanches are temporally dependent of one another, because they follow from the dynamics of the system.
Fig.~\ref{fig:avPS}B shows the power spectrum for this second approach averaged over many network disorder trials.
It is clear that the power spectrum follows $\Sps\sim1/f^b$ with $b\approx1.2$ for the critical case, $E=1.185\mV$.
The subcritical system also displays $\Sps\sim1/f^b$, but has $b\gtrsim1.5$ which does not correspond to long-range correlations.
The supercritical system has constant power spectrum,
corresponding to white noise (i.e. $1/f^b$ with $b=0$). In fact, within this approach, $\Sps(f)$ has been shown to display $1/f^b$ with
$b$ varying from $b\gtrsim1.5$ to $b=0$ while $E$ varies from $E=1.1\mV$ to $E=13\mV$~\cite{girardiV12015}.

\section{Concluding remarks}

We studied a model for the visual cortex of mammals consisting of a layered-columnar network of compartmental neurons.
The connections were randomly distributed according to structural parameters obtained from the literature~\cite{girardiV12015}.
This system presents an absorbing phase transition from an inactive phase to a percolating phase with the adjusted
parameters~\cite{girardiV12015}. The spreading of the signal is deterministic and advances one neuronal compartment per
time step. Thus, apparent inactive intervals between peaks of activity emerge because we measure only soma spikes.
This dynamics makes the avalanche definition ambiguous: one may either define an avalanche as all the activity sparked by a single
stimulus between consecutive absorbing states or as all the activity between two silent periods in the time series data.
We acknowledge that the second approach is more similar to the experimental approach for measuring local field potentials avalanches.
We use the EPSP parameter, $E$, in order to study the avalanche distributions in three different regimes: subcritical, critical
and supercritical.

Both approaches yield similar results in the critical region of the model: the avalanches are PL distributed both in size and duration with
similar exponents, contrary to what has been found for another model~\cite{hidalgo2017}. Even though the critical behavior is similar,
the avalanche distribution of both cases has different features with respect to the cutoff: while approach I yields a wider PL range for avalanche sizes and
duration, the PL from approach II presents accumulation of large events on the cutoff bump~\cite{pruessnerSOC2012}.
This highlights the strong dependence of the results on the model details.
The cutoff of these distributions define the dimensionality of the avalanches as expected.
However, the cutoffs of the distributions of silent period avalanches scale with the system size similarly to the expected for
critical systems even on the supercritical regime. Table~\ref{tab:exponents} shows the values of all the exponents of the model
for both approaches. The reported exponents have been verified via statistical tests yielding no more than 7\% error in their values.
Details of the tests may be found elsewhere~\cite{girardiV12015}.

\begin{table}[t!]
\caption{\label{tab:exponents}
Summary of the model fitted exponents for the absorbing state avalanches (approach I) and for the silent period avalanches (approach II).
Notice that the exponents defined here are not mutually exclusive because quantities have different interpretation in each approach.
For instance, the duration cutoff in approach I is simply related to the dimensionality of avalanches (because the total spreading
of the signal is constrained by the spatial extent of the network), whereas the duration cutoff in approach II is the autocorrelation
characteristic time~\cite{girardiV12015}. Moreover, $\tabs$ is related to the dynamical exponent inside the Griffiths phase and is
non-universal~\cite{juhaszEpidemic2015}. The exponents below have been verified through statistical tests and yielded no more than 7\% error in the
reported values. Details of the test may be found elsewhere~\cite{girardiV12015}.}
\begin{center}
\begin{tabular}{c|c|c|c}
\hline
\multicolumn{2}{c|}{\textbf{Quantity}} & \textbf{Approach I} & \textbf{Approach II}\\
\hline\hline
$\mathcal{P}(s)\sim s^{-\alpha}$ & Eq.~\eqref{eq:distS}      & 1.4(1) & 1.4(1)  \\
$\mathcal{P}(T)\sim T^{-\tau}$   & Eq.~\eqref{eq:distT}      & 1.4(1) & 1.5(1) \\
$s_c\sim L^{D}$                  & Eq.~\eqref{eq:cutoffS}    & 1.7(1) & 0.9(1) \\
$\bar{T}_c\sim L^{\bar{\mu}}$    & Eq.~\eqref{eq:cutoffT}    & 0.9(1) & -    \\
$T_c\sim \exp(kL^{D})$           & -                         & -      & 1.0(1) \\
$a=(\tau-1)/(\alpha-1)$          & -                         & 1.1(1) & 1.5(1) \\
$\mathcal{S}\sim f^{-b}$         & -                         & 0      & 1.2(1)    \\
\hline
\end{tabular}
\end{center}
\end{table}


Nevertheless, the results are significantly different in the subcritical and supercritical regimes: the absorbing state
avalanches do not present any PL distribution whereas the silent period avalanches presents a restricted PL shape for the distribution of
small avalanches both in size and duration (the power law does not scale with system size). This result is also dependent on model details,
given that other authors found no restricted power-laws outside of the critical range using another kind of model~\cite{scarpettaMem2013,scarpettaMem2014}.

The power spectra are also very different in the critical region of the model. This is expected due to the strong dependence
of the power spectrum on the avalanche definition. The independent avalanches generated by approach I results in a typical
white noise power spectrum, whereas approach II yields $1/f^b$ with $b\approx1.2$. These results are in agreement
with the power spectra of other models that rely on the inhibitory-excitatory synaptic balance mechanism~\cite{socPlasticity,lombardi1f2017}.

Absorbing state avalanches are more precise for finding the critical region of the model. However, there are no guarantees that an
experimental system will reach an absorbing state in between two measured avalanches. It is thus more likely that experimental
PL avalanche distributions that scale with system size alone will yield ambiguous results for defining the underlying system critical
behavior. Therefore, we emphasize that other quantities must be employed in order to rigorously define the critical region of the system.
For instance, we may define an order parameter-susceptibility pair for this model and show that the order parameter varies continuously
to zero whilst the susceptibility diverges as the critical point is approached~\cite{girardiV12015}. Such procedure for finding
the critical point is successfully applied in (non-)equilibrium statistical physics problems and is ultimately unambiguous.

\section*{Acknowledgments}

We thank discussions with M. Copelli, R. Dickman and G. S. Bortolotto. MGS thanks partial support from CNPq during the development of this work,
and funding from CIHR and the Mark Rayport and Shirley Ferguson Rayport Fellowship for Epilepsy Surgery of the Montreal Neurological
Institute and Hospital.


 \newcommand{\noop}[1]{}

\end{document}